\documentclass[12pt]{article}

\usepackage{amsmath,amssymb}

\newcommand{\e}{\varepsilon}
\newcommand{\p}{\bot}
\newcommand{\dd}{\partial}
\newcommand{\de}{\delta}
\newcommand{\m}{\mu}
\newcommand{\n}{\nu}
\newcommand{\ls}{\left(}
\newcommand{\rs}{\right)}
\newcommand{\po}{{\Pi_{\!\!\bot}}}
\newcommand{\tr}[1]{\overset{{\scriptscriptstyle 3}}{#1}{}}
\newcommand{\na}{\nabla\!}
\newcommand{\al}{\alpha}
\newcommand{\be}{\beta}
\newcommand{\g}{\gamma}
\newcommand{\ka}{\varkappa}
\newcommand{\sle}{\Rightarrow}
\newcommand{\vo}[1]{\underset{\thicksim}{#1}{}}
\newcommand{\pua}[2]{\left\{#1,#2\right\}}
\newcommand{\la}{\lambda}

\newcommand{\disn}[2]{$$\displaylines{\refstepcounter{equation}%
            \label{#1}\hskip 1em minus 1em #2\hfilneg}$$}
\newcommand{\nom}{\hfil\hskip 1em minus 1em (\theequation)}
\newcommand{\no}{\hfil \hskip 1em minus 1em\phantom{(\theequation)}%
            \hfilneg\cr\hfilneg\hskip 1em minus 1em\hfil}
\newcommand{\ns}{\hfill\cr\hfill}
\newcommand{\dis}[1]{$$\displaylines{#1}$$}

\makeatletter
\renewcommand{\section}{\@startsection{section}{1}{0pt}%
          {3.5ex plus 1ex minus .2ex}{2.3ex plus .2ex}{\noindent\hfil\bf}}
\makeatother

\begin{document}
\title{
Canonical formulation\\
of the embedded theory of gravity\\
equivalent to Einstein's General Relativity\\
}
\author{S.A.~Paston\thanks{E-mail: paston@pobox.spbu.ru},
V.A.~Franke\thanks{E-mail: franke@pobox.spbu.ru}\\
{\it St.-Petersburg State University, Russia}
}
\date{\vskip 15mm}
\maketitle

\begin{abstract}
We study the approach in which independent variables describing
gravity are functions of the space-time embedding into a flat
space of higher dimension. We formulate a canonical formalism for
such a theory in a form, which requires imposing additional constraints,
which are a part of Einstein's equations. As a result, we obtain a
theory with an eight-parameter gauge symmetry. This theory becomes
equivalent to Einstein's general relativity either after partial
gauge fixing or after rewriting the metric in the form that is
invariant under the additional gauge transformations. We write the
action for such a theory.
\end{abstract}

\newpage
\section{Introduction}
In the usual description of gravity in the framework of Einstein's
general relativity (GR), the four-dimensional space-time is a
Riemannian (more precisely, pseudo-Riemannian) space. An example
of the Riemannian space is a d-dimensional surface in a flat space
of a higher dimension if we consider the metric on this surface to
be induced by the trivial metric of the ambient space. But a
surface in a flat ambient space turns out to be not just a
particular example of a Riemannian space; considering an arbitrary
Riemannian space can be presumably replaced by considering such a
surface. According to the Janet-Cartan theorem \cite{gane,kart} (see,
e.~g., Remark 18 in \cite{kobno}), an arbitrary Riemannian space $W^d$ of
dimension $d$ can be locally embedded isometrically in any
Riemannian space of dimension greater than or equal to
 \disn{v3.1}{ N=\frac{d(d+1)}{2}, \nom} and therefore, in
particular, in a flat Riemannian space of such dimension. Friedman
\cite{frid} generalized this theorem to the case of a nonpositive
definite space signature.

This theorem ensures only a local
existence of an isometric embedding, i.~e., for a finite part of a
manifold. We note that if we address the problem of a global
manifold embedding, then the necessary dimension $N$ may increase
dramatically depending on the manifold topology (see Remark 18 in
\cite{kobno}).

The Janet-Cartan theorem ensures only the
existence, not the uniqueness, of the embedding. This means that
different surfaces with the same metric may exist. In this case,
we speak of a possible isometric bending of a surface. For
instance, it is clear that the uniqueness of the embedding is
certainly lost for $N>d(d+1)/2$ because we can first embed the
surface into a space of dimension $d(d+1)/2$ and then
isometrically embed this space in different ways as a part of a
cylinder in an $N$-dimensional space. Comparing the numbers of
variables and equations shows that in the general case, the
uniqueness of embedding presumably holds (up to trivial shifts and
rotations in the ambient space) when condition (\ref{v3.1}) is
satisfied exactly. Indeed, the metric $g_{\m\n}$ has $d(d+1)/2$
independent components, and we have the same number of equations
for $N$ functions describing the embedding. But in some cases, the
uniqueness may be absent even if condition (\ref{v3.1}) holds. As a
nontrivial example of a surface admitting isometric bending, we
can take a sufficiently small but finite part of a sphere.

In accordance with condition (\ref{v3.1}), we can take the ten-dimensional
space as an ambient space for the four-dimensional space-time. The
signature of the former can be conveniently taken to be
$(+,-,-,\dots,-)$, i.~e., we take the Minkowski space $R^{1,9}$ as an ambient
space. Because instead of considering space-time as a Riemannian
space, we can, with the above precaution, consider the
four-dimensional surface $W^4$ in the ten-dimensional space $R^{1,9}$,
the idea arises to use not the metric field $g_{\m\n}(x)$ but variables
describing the surface $W^4$ as independent variables in the gravity
description. As such variables, it is convenient to take the
embedding function $y^a(x^\m)$ describing the map
 \disn{v3.2}{
y^a(x^\m):R^4\longrightarrow R^{1,9}.
\nom}
Here and hereafter, the indices $a,b,\dots$ range the values
$0,1,2,\dots,9$ and ${\mu,\nu,\dots}=0,1,2,3$. We assume that $y^a$
are the Lorentzian coordinates in $R^{1,9}$ and we can raise and lower
the indices $a,b,\dots$ using the constant pseudo-Euclidean metric
of the ambient space
\dis{
\eta_{ab}=diag(1,-1,-1,\dots,-1).
}

This idea was
first advanced in 1975 in a talk by Regge and Teitelboim and was
then published in \cite{regge}. In this way, we can try to study another
approach to quantizing gravity and, in particular, obtain a new
insight into the causality problem in the theory because we have
fixed light cones in the ambient flat space. Moreover, we can
relate this theory to other theories, such as superstring theory,
using the extra dimensions.

In \cite{regge}, the theory action was chosen
to be the standard GR action
 \disn{39}{
S=\int d^4x\, \sqrt{-g}\;R,
\nom}
where $R$ is the
scalar curvature and $g=\det g_{\m\n}$, where the metric is induced and
expressed in terms of the embedding function,
 \disn{nn1}{
g_{\m\n}=\dd_\m y^a\, \dd_\n y_a.
\nom}
Varying this action with respect to $y^a(x)$ produces the equations
of the theory of embedding:
 \disn{nn2}{
\na_\m\ls G^{\m\n}\dd_\n y^a\rs=0,
\nom}
where $G^{\m\n}$ is
Einstein's tensor constructed from metric (\ref{nn1}) and $\na_\m$ is the
covariant derivative induced by this metric. In the case of matter
with the energy-momentum tensor $T^{\m\n}$, Eqs.~(\ref{nn2})
become
 \disn{nn3}{
\na_\m\ls\ls G^{\m\n}-\ka\,T^{\m\n}\rs\dd_\n y^a\rs=0.
\nom}
Because adding matter does not play a principal
role in describing the theory, for simplicity in what follows, we
consider the gravitational field with matter absent.

Equations (\ref{nn2})
are more general than Einstein's equations, i.~e., any solution of
Einstein's equations is a solution of the theory of embedding, but
not vice versa. The theory of embedding is therefore not
equivalent to GR. There are extra solutions in the former. The
question therefore arises whether it is possible to introduce
additional restrictions into the theory of embedding such that it
becomes equivalent to GR.

To exclude extra solutions, the
following idea was advanced in \cite{regge}: it was proposed to complete
the set of equations of motion arising from the action by imposing
additional constraints $G_{\m\p}=0$ on some of Einstein's equations,
where the symbol $\p$ denotes the direction orthogonal to the
constant-time surface. We discuss this possibility in Sec.~3
below, where we show that at least in the general case, imposing
the constraints $G_{\m\p}=0$ only at the initial instant suffices to
ensure the equivalence to Einstein's equations, i.~e., these
constraints are analogous to the first-class constraints in the
canonical formalism. This result was derived in more detail in
\cite{sbshk05}, where a detailed exposition of the formalism convenient for
describing the theory of embedding can also be found. The basic
equations of this formalism are given in Sec.~2.

We mention that
an artificial, {\it ad hoc}, introduction of additional equations into
the theory seems not quite satisfactory, as was noted in \cite{deser},
where it was also argued that a proper way out might be to find a
modification of the action that generates the necessary additional
equations, but it was also mentioned that how to do this is
unknown. We propose such a modification of the action in Sec.~5
below.

There were attempts to construct a canonical formalism for
the theory of embedding. It is known that disregarding surface
terms allows reducing the gravitational action to a form in which
the Lagrangian contains only first derivatives of the metric $g_{\m\n}$
with respect to time. Because formula (\ref{nn1}) contains the
differentiation, the Lagrangian turns out to contain second-order
time derivatives of the embedding function $y^a$. This considerably
hinders constructing the canonical formalism. Nevertheless, a
possibility of constructing a canonical formulation of this theory
was studied in \cite{tapia} based on a special technique developed for this
case. The progress along this route was achieved only after a
complete gauge fixing.

A detailed consideration shows that the
embedding theory equations do not contain time derivatives of $y^a$
of order higher than two. This would indicate that the action can
be rewritten in a form in which the Lagrangian contains only first
derivatives in time. It was already noted in \cite{regge} that this can be
achieved if the action is written in the Arnowitt-Deser-Misner
form \cite{adm}. If we represent the action in this form, then we can
develop the canonical formalism standardly, but extremely complex
constraints appear in the theory. Their form was studied in \cite{frtap},
where it was found that some of the constraints cannot be written
explicitly and can be represented only in the form of the
existence of coincident roots of a certain pair of polynomials.

A possibility of adding the conditions $G_{\m\p}=0$ (which can also be
considered constraints) to the set of arising constraints was
considered in \cite{regge}, but the problem of closing such a unified
system of constraints remained open. We devote Sec.~4 of this
paper to constructing such a canonical formalism with the
additional constraints. The found system of constraints, which is
found explicitly in this case, differs from that in \cite{regge} because
the restrictions on the generalized momentums that appeared there
were not taken into account correctly. We demonstrate that the
obtained system of constraints is closed.

In Sec.~5, we construct
and analyze the action corresponding to the found canonical
formulation of the theory.

An extended bibliography related to the
theory of embedding and related questions can be found in \cite{tapiaob}.

\section{A concise exposition of the embedding theory formalism}
The description of a surface $W^4$ in the flat space $R^{1,9}$ using the
embedding function $y^a(x)$ is invariant under transformations of the
coordinates $x^\m$ on the surface. The components of the functions
$y^a(x)$ act like scalars with respect to these transformations. We
can therefore regard the function $y^a(x)$ as a ten-component field
defined in the four-dimensional Riemann space and carrying the
superscript of the global internal symmetry group $SO(1,9)$,
corresponding to the Lorentz transformations of the ambient space
$R^{1,9}$.

Because $y^a(x)$ is a scalar, its covariant derivative
coincides with the standard derivative,
 \disn{1}{
\na_\m\, y^a=\dd_\m y^a\equiv e^a_\m.
\nom}
Analogously, the covariant derivatives of quantities carrying both
Greek and Latin indices are constructed standardly as if there
were no Latin indices. The quantity $e^a_\m$ resembles a tetrad used
in the tetrad description of gravity, but it differs from the
standard tetrad in that its index $a$ ranges more values than the
index $\m$. This quantity can be treated as the union of four vectors
(if we take $\m=0,1,2,3$) of the ambient space. These vectors
constitute a basis (nonorthogonal in general) in the subspace
tangent to the surface $W^4$ at a given point. At the same time, the
quantity $e^a_\m$ is a vector with respect to its index $\m$ at a fixed
$a$.

The induced metric is
 \disn{2}{
g_{\m\n}=e^a_\m e^b_\n\,\eta_{ab}=e^a_\m \, e_{\n,a}=
\dd_\m y^a\, \dd_\n y_a.
\nom}
It is convenient to introduce the quantity
 \disn{10.1}{
e^\m_a=g^{\m\n}e_{\n,a},\qquad
g^{\m\n}g_{\n\al}=\de^\m_\al.
\nom}
It is then easy to note that the equalities
 \disn{10.3}{
e^\g_a e^a_\be=\de^\g_\be,\qquad
g^{\m\n}=e^{\m,a}e^\n_a
\nom}
are satisfied. From the
condition of covariant constancy of the metric and the absence of
torsion (because of which $\na_\m e^a_\n=\na_\n e^a_\m$)
and from form (\ref{2}) of the
metric, we can establish (see \cite{sbshk05}) that
 \disn{10}{
e_{\al,a}\na_\m e^a_\n=0,
\nom}
whence we expression the connection
 \disn{14}{
\Gamma^\be_{\m\n}=e^\be_a\;\dd_\m e^a_\n=e^\be_a\;\dd_\m\dd_\n y^a.
\nom}
But in calculations within this formalism, we can disregard
such a noncovariant quantity as the connection, which is a
definite advantage of the formalism. Usually, we cannot write a
formula for the covariant differentiation without a connection,
but it is possible here. We write the covariant derivative of a
vector using property (\ref{10}):
 \disn{10.2}{
\na_\al a^\m=e^\m_a e^a_\n\na_\al a^\n=
e^\m_a\na_\al(e^a_\n a^\n)-e^\m_a(\na_\al e^a_\n) a^\n=
e^\m_a\na_\al(e^a_\n a^\n)=
e^\m_a\,\dd_\al(e^a_\n a^\n).
\nom}
We can
write analogous formulas for the covariant differentiation of
tensors of arbitrary rank. We obtain a simple rule for the
covariant differentiation: contracting with respect to every index
with the quantity $e^a_\n$, we must "transfer" the tensor from the
Riemannian space to the ambient space, take the standard
derivative there, and then "transfer" it back performing the
contraction with $e^\n_a$. In cases where it cannot lead to confusion,
we merely write $a^a$ instead of the contraction $e^a_\n a^\n$ (and
analogously for quantities with several indices).

We now introduce
the quantity $\Pi^a_b(x)$, which is extremely useful for calculations
and is the projection on the plane tangent to the surface $W^4$ at a
given point. It is easy to verify that such a projection operator
can be written as
 \disn{16}{
\Pi^a_b=e^a_\m e^\m_b.
\nom}
It is also convenient to
introduce the operator of projection to the space dual to the
tangent plane: \disn{16.1}{
\po^a_b=\de^a_b-\Pi^a_b.
\nom}
We write several
properties of the introduced objects useful for calculations (see
the proof in \cite{sbshk05}):
 \disn{p1}{
\Pi_{ab}=\Pi_{ba},\qquad
\de \Pi_{ab}=-\de\po_{ab},\qquad
\Pi^a_b(\de\Pi^b_c)\Pi^c_d=0,\no
\po^a_b(\de\Pi^b_c)\po^c_d=0,\qquad
\de\Pi_{ab}=\Pi^c_a(\de\Pi_{cd})\po^d_b+\po^c_a(\de\Pi_{cd})\Pi^d_b.
\nom}

The second fundamental form of the surface $b^a_{\m\n}$ plays
an important role in describing the geometry of embedded surfaces.
By definition (see, e.~g., Sec.~3 in Chap.~7 in \cite{kobno}), for any
tangent vector field $f^b$,
 \disn{18.1}{
(\dd_\m f^b)\po^a_b=b^a_{\m\n}f^\n,
\nom}
whence,
noting that $f^b\po^a_b=0$, we obtain the equality
 \disn{22}{
b^a_{\m\n}=e^b_\n\,\dd_\m\Pi^a_b.
\nom}
We note that the condition  \disn{22.1}{
\Pi^b_a\, b^a_{\m\n}=0
\nom}
is satisfied
identically for $b^a_{\m\n}$, i.~e., $b^a_{\m\n}$
can be regarded as the set of
vectors indexed by $\m$ and $\n$ with components indexed by $a$ that are
orthogonal to the tangent plane. We have one more convenient
representation for the second fundamental form of the surface.
Replacing $\Pi^a_b$ with $-\po^a_b$ in (\ref{22})
and pushing the derivative to $e^b_\n$,
we obtain the equality
 \disn{22.2}{
b^a_{\m\n}=\po^a_b\,\dd_\m e^b_\n=\po^a_b\,\dd_\m\dd_\n y^b,
\nom}
whence we immediately see that $b^a_{\m\n}$ is symmetric in the
lower indices. We can also easily note that formula (\ref{22.2}) can be
rewritten in the explicitly covariant form
 \disn{21}{
b^a_{\m\n}=\na_\m e^a_\n=\na_\m \na_\n\, y^a.
\nom}

In the case where the codimension of the surface is one and
hence $\po_{ab}=n_a n_b$, because of condition (\ref{22.1}),
instead of $b^a_{\m\n}$, it
suffices to consider the quantity
 \disn{24.1}{
K_{\m\n}=n_a b^a_{\m\n},
\nom}
which is
also called the second fundamental (or second quadratic) form of
the surface in this case.

The second fundamental form of the
surface plays an important role in describing gravity in terms of
the embedding function because the Riemann-Christoffel curvature
tensor in the case of a flat ambient space is expressed in terms
of precisely this quantity:
 \disn{27.1}{
R_{\al\be\m\n}=b^a_{\al\m} b_{a,\be\n}-b^a_{\al\n} b_{a,\be\m}.
\nom}
We note that this equation is the Gauss equation for a surface
embedded in a flat ambient space. The scalar curvature can be
written in the form
 \disn{31}{
R=g^{\al\m}g^{\be\n}R_{\al\be\m\n}=
\ls g^{\al\m}g^{\be\n}-g^{\al\n}g^{\be\m}\rs
b^a_{\al\m} b_{a,\be\n}.
\nom}

Substituting expression (\ref{27.1}) in the known representation of
the Einstein tensor,
 \disn{36}{
G^{\m\n}\equiv R^{\m\n}-\frac{1}{2}\,g^{\m\n}R=
\frac{1}{4}\,g_{\xi\zeta}E^{\m\xi\al\be}E^{\n\zeta\g\de}R_{\al\be\g\de},
\nom}
where $E^{\m\xi\al\be}=\frac{1}{\sqrt{-g}}\,\e^{\m\xi\al\be}$
and $\e^{\m\xi\al\be}$ is the
unit totally antisymmetric tensor, we can write it in the form
 \disn{36.1}{
G^{\m\n}=
\frac{1}{2}\,g_{\xi\zeta}E^{\m\xi\al\be}E^{\n\zeta\g\de}\,
b^a_{\al\g} b_{a,\be\de}.
\nom}
We use this expression for
the Einstein tensor below.

\section{Comparing equations of the
embedding theory with Einstein's equations}
Embedding theory
equations (\ref{nn2}) are ten equations obtained by varying with respect
to the ten components of the embedding function $y^a$. But among
these equations, four are satisfied identically. Indeed, Eqs.~(\ref{nn2})
can be rewritten as the set of equations
 \disn{48}{
\Pi^b_a\,\na_\m\ls G^{\m\n}e^a_\n\rs=0,
\nom}
\vskip -2em
 \disn{48.1}{
\po^b_a\,\na_\m\ls G^{\m\n}e^a_\n\rs=0.
\nom}
By virtue of formulas (\ref{21}),(\ref{22.1})
the first of these equations can be written equivalently in the
form
 \disn{48.2}{
\na_\m G^{\m\n}=0
\nom}
and is satisfied identically because of the
Bianci identity. Therefore, only six independent embedding
equations (\ref{48.1}) remain. Again using (\ref{21})
and taking identity (\ref{48.2})
into account, we can rewrite them in the form
 \disn{50}{
G^{\m\n}\,b^a_{\m\n}=0.
\nom}
We recall that the quantity $b^a_{\m\n}$ is regarded as the set of
vectors with components indexed by $a$ directed normal to the
tangent plane, and Eqs.~(\ref{50}) therefore contain only six
independent equations.

We now rewrite Eq.~(\ref{50}) using formula (\ref{36.1}):
 \disn{50.1}{
g_{\xi\zeta}E^{\m\xi\al\be}E^{\n\zeta\g\de}\,
b^b_{\al\g}\: b_{b,\be\de}\: b^a_{\m\n}=0.
\nom}
It is clear in this form of
writing that the embedding theory equations do not contain
derivatives of an order higher than two. It is interesting that
Einstein's equations have the same property if they are written in
terms of $y^a$:
 \disn{52}{
g_{\xi\zeta}E^{\m\xi\al\be}E^{\n\zeta\g\de}\,
b^b_{\al\g}\: b_{b,\be\de}=0.
\nom}

We now compare the
embedding theory equations and Einstein's equations. For this, we
fix the choice of time on the surface $W^4$, i.~e., we define a
system of three-dimensional spacelike surfaces $W^3$ corresponding to
constant time. These surfaces are described by the embedding
functions
 \disn{n0}{
y^a(x^i)\equiv y^a(x^\m)|_{x^0=t}.
\nom}
Here and hereafter, the
indices $i,k,\dots$ range the values $1,2,3$. For each such
surface, we can introduce all the quantities described in Sec.~2,
and we label such quantities with the digit 3 over the letter:
$\tr{e}^a_i,\; \tr{g}_{ik},\; \tr{\Pi}_{ab},\; \tr{\po}_{ab},\;
\tr{E}_{abc},\;
\tr{b}^a_{ik},\; \tr{R}_{iklm},\dots$.
We note that tensors with upper Latin indices marked by the digit 3
can be always obtained by raising these indices using the matrix
$\tr{g}^{ik}$, which is inverse to the matrix $\tr{g}_{ik}$.
We also introduce the
unit vector $n_a$ that is tangent to the surface $W^4$ at a given point
and is normal to $W^3$. From the first equality in (\ref{10.3}),
we have
 \disn{n2}{
e^0_a\; e^a_i=0 \quad\sle\quad e^0_a\;\tr{e}^a_i=0,
\nom}
whence we obtain
 \disn{n3}{
n_a=\frac{e^0_a}{\sqrt{e^0_b\; e^{0,b}}}=\frac{e^0_a}{\sqrt{g^{00}}}.
\nom}
For definiteness, we set $n_0>0$. It
is clear that
 \disn{n3.1}{
\Pi_{ab}=\tr{\Pi}_{ab}+n_an_b,\qquad
\po_{ab}=\tr{\po}_{ab}-n_an_b.
\nom}
We obtain one more useful representation for the vector $n^a$.
Using formulas (\ref{n3.1}),(\ref{n3}), we have
 \disn{n3.1a}{
\tr{\po}^a_be^b_0=n^an_be^b_0=\frac{n^ae^0_be^b_0}{\sqrt{g^{00}}}=
\frac{n^a}{\sqrt{g^{00}}},
\nom}
whence
 \disn{n3.1b}{
n^a=\sqrt{g^{00}}\,\,\tr{\po}^a_b\,\dd_0y^b=
\frac{\tr{\po}^a_b\,\dd_0y^b}{\sqrt{\dd_0y^c\,\tr{\po}_{cd}\,\dd_0y^d}}.
\nom}

As is known, the second fundamental
form of the surface $W^3$ as a submanifold in $W^4$ is given by
 \disn{n6}{
K_{ik}=-\na_i\,  n_k,
\nom}
where the covariant derivative is determined by the
Riemannian connection in $W^4$. Using covariant differentiation rule
(\ref{10.2}), we find that
 \disn{n6.1}{
K_{ik}=-e^a_k\,\dd_i(e^\m_a n_\m)=-\tr{e}^a_k\,\dd_i n_a=
n_a\,\dd_i\tr{e}^a_k=n_a\,\tr{b}^a_{ik}=n_a\,\dd_i\dd_k y^a.
\nom}
Applying formula (\ref{22.2}) for the
second fundamental forms of the surfaces $W^4$ and $W^3$ and using the
second relation in (\ref{n3.1}), we can easily obtain the simple law for
adding second fundamental forms:
 \disn{n10}{
b^a_{ik}=\tr{b}^a_{ik}-n^aK_{ik}.
\nom}

Using the above relations, among all Einstein's equations taken in
form (\ref{52}), those four equation that can be written as
 \disn{n16}{
n_a G^{ac}=0,
\nom}
can be represented as the equation
 \disn{103.2}{
\ls \tr{g}^{ik}\tr{g}^{lm}-\tr{g}^{il}\tr{g}^{km}\rs
\tr{b}^a_{ik}\,\tr{b}^b_{lm}\,n_a n_b
-\tr{R}=0,
\nom}
taken together with the three
equations
 \disn{110}{
\ls \tr{g}^{ik}\tr{g}^{lm}-\tr{g}^{im}\tr{g}^{lk}\rs
\,\tr{b}^a_{im}\,\dd_l n_a=0
\nom}
(see \cite{sbshk05}). Applying the identity
\disn{nn4}{
\tr{\po}^b_a \tr{\na}_l\ls
\ls \tr{g}^{ik}\tr{g}^{lm}-\tr{g}^{im}\tr{g}^{lk}\rs
\,\tr{b}_{a,im}\rs=
\tr{\po}^b_a\, \tr{g}^{ik}\tr{g}^{lm}\ls
\tr{\na}_l\tr{\na}_i \tr{e}^a_m-\tr{\na}_i\tr{\na}_l \tr{e}^a_m\rs=\ns=
\tr{\po}^b_a\,\tr{g}^{ik}\tr{g}^{lm}
\tr{R}^n{}_{mli}\tr{e}^a_n=0,
\nom}
where we use formula (\ref{21}),
we can conveniently transform Eqs.~(\ref{110}) to
the form
\disn{nn5}{
\tr{\na}_l\ls
\ls \tr{g}^{ik}\tr{g}^{lm}-\tr{g}^{im}\tr{g}^{lk}\rs
\,\tr{b}^a_{im}\, n_a\rs=0.
\nom}

We can easily see that Eqs.~(\ref{103.2}) and (\ref{nn5})
do not contain time
derivatives of $y^a(x)$ of an order higher than one, and such
derivatives enter the equations only through the quantity $n^a$. Four
equations (\ref{n16}) can be therefore considered constraints, i.~e.,
conditions, which impose restrictions on the initial data $y^a(x^i)$,
$\dd_0y^a(x^i)$ defined on the surface $x^0=const$. It is interesting that
Eqs.~(\ref{n16}), which were constraints in GR, remain constraints in the
embedding theory although the GR independent variables (the metric
$g_{\m\n}$) are now expressed in terms of the independent variables of
the embedding theory (the functions $y^a(x)$) by means of
differentiation. We call Eqs.~(\ref{n16}) Einstein's constraints.

We now
consider the previously obtained embedding theory equations (\ref{50}),
writing them in the form
 \disn{n16.1}{
G^{ab}\,b^c_{ab}=0.
\nom}
We assume that they
are satisfied and treat them as the equations describing the time
evolution of the three-dimensional spacelike surface $W^3$. We find
what additional restrictions must be introduced in the theory for
it to be equivalent to GR, i.~e., for Einstein's equations to
hold. The analysis just performed demonstrates that for this, we
must at least choose the initial data, i.~e., the values of $y^a(x^i)$
and $\dd_0y^a(x^i)$ at the initial instant, that satisfy Einstein's
constraints.

We assume that this was done, i.~e., Eqs.~(\ref{n16}) are
satisfied at the initial instant. Using formula (\ref{n3.1}) at the
initial instant, we can then write the expression
 \disn{n18.1}{
G^{ab}\,\tr{e}^i_a\tr{e}^k_b\;b^h_{ik}=0.
\nom}
instead of (\ref{n16.1}). The quantity $b^h_{ik}$ in this
equation can be interpreted as a matrix with the multi-indices $h$
and $\{ik\}$. We can then assume that $h$ ranges not ten but six values
because four identities (\ref{22.1}) are satisfied, and the multi-index
$\{ik\}$ also ranges six values because the quantity $b^h_{ik}$ is
symmetric. The quantity $b^h_{ik}$ can therefore be considered a square
matrix of the size $6\times 6$. We additionally assume that this matrix
is nondegenerate at all points of the initial surface, which can
be conditionally written as
 \disn{n18.2}{
\det\ls b^h_{ik}\rs\ne 0.
\nom}
This assumption
is purely technical; it just excludes a certain initial data
subset of measure zero. Condition (\ref{n18.2}) was discussed
in \cite{sbshk05}. In
particular, it was shown there that breaking this condition
results in a special situation for Einstein's equations written in
terms of the embedding function.

If the matrix $b^h_{ik}$ is
nondegenerate, then Eq.~(\ref{n18.1}) is equivalent to the equation
 \disn{n19}{
G^{ab}\,\tr{e}^i_a\tr{e}^k_b=0,
\nom}
which, together with imposed constraints (\ref{n16}),
can be written as Einstein's equations
 \disn{n19.1}{
G^{\m\n}=0.
\nom}
We have thus
obtained the first result: the embedding theory equations together
with constraints (\ref{n16}) and condition (\ref{n18.2}),
imposed at some instant,
result in Einstein's equations being satisfied at this instant.

It was shown in \cite{sbshk05}
that using the Bianci identity, we can obtain the
relation for the time derivative of constraints (\ref{n16}):
 \disn{n24}{
\dd_0\ls n_a G^{ac}\rs=
-\frac{1}{\sqrt{g^{00}}}\,\Pi^c_b e^i_a\; \dd_i G^{ab}-
n_a G^{ab}\,\dd_0\po^c_b+G^{ac}\dd_0n_a.
\nom}
This relation easily implies the second result: if Einstein's equations
(\ref{n19.1}) are satisfied at some instant, then the time derivative of
constraints (\ref{n16}) vanishes at this instant.

Together with the
first result above, we conclude that if the embedding theory
equations are always satisfied, condition (\ref{n18.2}) is satisfied during
some time interval, and Einstein's constraints (\ref{n16}) are
imposed at the beginning of this time interval, then Einstein's
equations are satisfied during the whole time interval. Because,
in particular, Einstein's constraints, being imposed at the
initial instant, are satisfied automatically at subsequent times,
we can expect that these constraints become first-class
constraints in the framework of a canonical formalism.

If the
initial data are chosen as corresponding to the general case and
condition (\ref{n18.2}) is therefore satisfied, then the embedding theory
with Einstein's constraints imposed at the initial instant is
equivalent to GR.

\section{Canonical formalism with additionally imposed
Einstein's constraints}
We now develop the canonical formalism for
the embedding theory. Dropping the total divergence term in the
integrand in action (\ref{39}), we write it in the Arnowitt-Deser-Misner
form
 \disn{77.1}{
S=\int d^4x\, \sqrt{-g}\ls(K^i_i)^2-K_{ik} K^{ik}+\tr{R}\rs.
\nom}
If we rewrite
this expression in terms of the embedding function $y^a(x)$ and use
one of the forms of representing formula (\ref{n6.1}),
then it becomes  \disn{77.2}{
S=\int d^4x \sqrt{-g}\ls
n_a\, n_b\;
\tr{b}^a_{ik}\,\tr{b}^b_{lm}
L^{ik,lm}+\tr{R}\rs,
\nom}
where
\disn{t7}{
L^{ik,lm}=\tr{g}^{ik}\tr{g}^{lm}-\tr{g}^{il}\tr{g}^{km}.
\nom}
We note that formula (\ref{n3.1b}) implies the
equality
\disn{t0}{
g^{00}=\frac{1}{\dot y^a\;\tr{\po}_{ab}\;\dot y^b},
\nom}
where $\dot y^a\equiv \dd_0 y^a$. Using the
relation $g=\tr{g}/g^{00}$ and
formulas (\ref{n3.1b}),(\ref{t0}),(\ref{27.1}), we can
rewrite action (\ref{77.2}) in the form in which the derivatives of the
variables $y^a(x)$ with respect to the time $x^0$ are written
explicitly:
\disn{t1}{
S=\int dx^0\, L(y^a,\dot y^a),\qquad
L=\int d^3x\;\frac{1}{2}\ls
\frac{\dot y^a\;B_{ab}\;\dot y^b}{\sqrt{\dot y^a\;\tr{\po}_{ab}\;\dot y^b}}+
\sqrt{\dot y^a\;\tr{\po}_{ab}\;\dot y^b}\;B^c_c\rs,
\nom}
where the quantity
\disn{t2}{
B^{ab}=2\sqrt{-\tr{g}}\;\tr{b}^a_{ik}\tr{b}^b_{lm}L^{ik,lm},
\nom}
and also the projection operator $\tr{\po}_{ab}$
do not contain time derivatives.

We can treat the quantity $\tr{b}^a_{ik}$
as the set of six vectors (at the fixed values of the indices
$i$ and $k$ with respect to which it is symmetric). On the other hand,
this quantity satisfies three identities $\tr{b}^a_{ik}\tr{e}_{a,l}=0$.
In the general case, we therefore have a unique normalized vector $w_a$
determined by the conditions
\disn{t3}{
w_a\tr{e}^a_l=0,\qquad w_a\tr{b}^a_{ik}=0.
\nom}
We note that action of the matrix $B^{ab}$ on this vector gives
zero, and this matrix is therefore not invertible even in the
seven-dimensional space orthogonal to the vectors $\tr{e}^a_i$. This was
not mentioned in \cite{regge}, which resulted in an incorrect form of one
of the constraints.

We find the generalized momentum $\pi_a$ for the
variable $y^a$ from action (\ref{t1}) (we use
formulas (\ref{n3.1b}),(\ref{t0})):
\disn{t10}{
\pi_a=\frac{\de L}{\de \dot y^a}=
B_{ab}n^b-\frac{1}{2}n_a\ls n_c B^{cd} n_d-B^c_c\rs.
\nom}
Taking the
properties of the quantity $\tr{b}^a_{ik}$ into account, we obtain the
constraints
\disn{t12}{
\Phi_i=\pi_a\tr{e}^a_i=0.
\nom}
In the general case, relation
(\ref{t10}) must generate one more constraint arising as a restriction on
the momentum $\pi_a$ following from the identity $n^an_a=1$. We note that
it is extremely difficult to write this constraint as an algebraic
expression (this problem was studied in \cite{frtap}). But instead of
studying the general case, we assume that Einstein's constraints
(\ref{n16}) are additionally imposed when constructing the canonical
formalism. Using formulas
(\ref{103.2}),(\ref{nn5}),(\ref{t2}),(\ref{27.1}), we write
these constraints in the form
\disn{t16}{
{\cal H}^i=-2\sqrt{-\tr{g}}\;\,\tr{\na}_k\ls L^{ik,lm}
\,\tr{b}^a_{lm}\, n_a\rs=0,
\nom}
\vskip -1em
\disn{t13}{
{\cal H}^0= n_c B^{cd} n_d-B^c_c=0,
\nom}
where we choose the
common factors for convenience. Accounting for constraint (\ref{t13}) in
expression (\ref{t10}) for the momentum, we obtain the momentum in the
form
\disn{t17}{
\pi_a=B_{ab}n^b.
\nom}
As a result, the identity $n^an_a=1$ does not
restrict the momentum $\pi_a$ (because the matrix $B_{ab}$, as stated above,
is not invertible even in the subspace orthogonal to the vectors
$\tr{e}^a_i$); instead, the constraint
\disn{t18}{
\Phi_0=\pi_a w^a=0
\nom}
arises in
addition to constraint (\ref{t12}).

Using formulas (\ref{t10}),(\ref{t1}),(\ref{n3.1b}),(\ref{t0}),
we can easily find that the theory Hamiltonian
\disn{t18.1}{
H=\int\! d^3x\, \pi_a \dot y^a - L
\nom}
vanishes. The generalized Hamiltonian then
reduces to a linear combination
of constraints (\ref{t12})-(\ref{t13}),(\ref{t18}),
and we must verify whether these constraints are governed by
a constraint algebra of the first kind.

In the canonical
formalism, constraints must be expressed via generalized
coordinates and momenta, i.~e., via $y^a$ and $\pi_a$
but not $\dot y^a$ in our case. Constraints (\ref{t12}) and (\ref{t18})
satisfy this requirement (we note
that the vector $w_a$ determined by conditions (\ref{t3}) depends
on $y^a$ but
not on $\dot y^a$), while constraints (\ref{t16}) and (\ref{t13})
do not satisfy it.
They must therefore be transformed to the necessary form. For
this, we introduce the quantity $\al^{ik}_a$ unambiguously determined by
the conditions
\disn{t20}{
\al^{ik}_a=\al^{ki}_a,\quad
\al^{ik}_a\, \tr{e}^a_l=0,\quad
\al^{ik}_a w^a=0,\quad
\al^{ik}_a\, \tr{b}^a_{lm}=\frac{1}{2}\ls\de^i_l\de^k_m+\de^i_m\de^k_l\rs.
\nom}
It is clear that
this quantity as well as $w_a$ depends on $y^a$ but not on $\dot y^a$.
Relation
(\ref{t17}) implies that
\disn{t22}{
\tr{b}^b_{ik}n_b=\frac{1}{2\sqrt{-\tr{g}}}\;
\hat L_{ik,lm}\,\al^{lm}_a\,\pi^a,
\nom}
where
\disn{t21}{
\hat L_{pr,lm}=\frac{1}{2}\,g_{pr}g_{lm}-g_{pl}g_{rm},\qquad
\hat L_{pr,lm}L^{ik,lm}=\de^i_p\,\de^k_r.
\nom}
Using formula (\ref{t22}), we can write
constraints (\ref{t16}),(\ref{t13}) as
\disn{t27}{
{\cal H}^i=-\sqrt{-\tr{g}}\;\tr{\na}_k\!\ls
\frac{1}{\sqrt{-\tr{g}}}\;\pi^a\al_a^{ik}\rs,
\nom}
\vskip -1em
\disn{t28}{
{\cal H}^0=\frac{1}{2\sqrt{-\tr{g}}}\;\pi^a\al_a^{ik}\hat L_{ik,lm}\al^{lm}_b\pi^b-
2\sqrt{-\tr{g}}\;
\tr{R}.
\nom}
Constraints
represented in this form cease to contain $\dot y^a$. It is clear that
constraint (\ref{t28}) is an expression quadratic in the momentum $\pi^a$ and
constraint (\ref{t27}) and also constraints (\ref{t12}),(\ref{t18})
are linear in
this momentum.

We note that the set of constraints
(\ref{t12}),(\ref{t18}),(\ref{t27}),(\ref{t28}) found
here differs from that in \cite{regge} in that
constraint (\ref{t18}) looks different (see the note
after formula (\ref{t3})).

We must now calculate the Poisson brackets between obtained
constraints (\ref{t12}),(\ref{t18}),(\ref{t27}), (\ref{t28}).
Because Einstein's
constraints, as shown in Sec.~3, must be preserved on the
equations of motion, it can be hoped that these Poisson brackets
are expressed by linear combinations of the constraints, i.~e., we
have a first-class constraint algebra. But this must be verified
explicitly.

It is useful to find the result of the action of
transformations generated by constraints
(\ref{t12}),(\ref{t18}),(\ref{t27}) on
some combinations of variables. Calculation shows that
\disn{t51}{
\pua{\int\! d^3\tilde x\, \vo{\Phi}_i\vo{\xi}^i}{y^a}=
\xi^i\dd_i y^a,\qquad
\pua{\int\! d^3\tilde x\, \vo{\Phi}_i\vo{\xi}^i}{\frac{\pi_a}{\sqrt{-\tr{g}}}}=
\xi^i\dd_i \frac{\pi_a}{\sqrt{-\tr{g}}},
\nom}
\vskip -1em
\disn{t68}{
\pua{\int\! d^3\tilde x\, \vo{\Phi}_0\vo{\xi}}{\tr{g}_{lm}}=0,\qquad
\pua{\int\! d^3\tilde x\, \vo{\Phi}_0\vo{\xi}}{\pi^a\al_a^{lm}}\approx 0,
\nom}
\vskip -1em
\disn{t62}{
\pua{\int\! d^3\tilde x\, \ls \vo{\cal H}_i+\vo{\Phi}_i\rs\vo{\xi}^i}{\tr{g}_{lm}}=0,\qquad
\pua{\int\! d^3\tilde x\, \ls \vo{\cal H}_i+\vo{\Phi}_i\rs\vo{\xi}^i}{\pi^a\al_a^{lm}}\approx 0,
\nom}
where ${\cal H}_i=\tr{g}_{ik}{\cal H}^k$ and $\xi,\xi^i$ are
arbitrary infinitesimal quantities depending on $x^k$, $\{\dots\}$ is
the Poisson bracket, the sign $\approx$ denotes the equality up to adding
a linear combination of constraints, and we use the notation
$f\equiv f(x)$, $\vo{f}\equiv f(\tilde x)$.

Formulas (\ref{t51}) imply that the constraint $\Phi_k$ generates
transformations of three-dimensional coordinates on the
constant-time surface $W^3$ (we note that the generalized momentum $\pi^a$
is a three-dimensional scalar density). Because the constraints $\Phi_0$
and ${\cal H}^0$ are scalar densities and $\Phi_i$ and ${\cal H}^i$
are vector densities,
this means that the Poisson brackets between the constraint $\Phi_i$ and
all other constraints are linear combinations of the constraints.

The first formula in (\ref{t68}) states that the constraint $\Phi_0$
generates
a transformation that is an isometric bending of the surface $W^3$.
Because the constraints ${\cal H}^i$ and ${\cal H}^0$
are expressed via the quantities
$\tr{g}_{lm}$ and $\pi^a\al_a^{lm}$, formulas (\ref{t68})
imply that the Poisson brackets between the constraint $\Phi_0$ and the
constraints ${\cal H}^i$, ${\cal H}^0$ are linear combinations
of the constraints.

Analogously, formulas (\ref{t62}) demonstrate that the constraint
combination ${\cal H}_i+\Phi_i$ also generates isometric bendings of the
surface $W^3$ and that its Poisson brackets with the constraints
${\cal H}^i$, ${\cal H}^0$
reduce to linear combinations of the constraints. We note
that the total number (four) of the found generators of
three-dimensional isometric bendings corresponds to comparing the
number of independent components of the three-dimensional metric
(six) and the dimension (ten) of the space into which the
three-dimensional surface is embedded.

Taking all the above into
account, to prove the closedness of the constraint algebra
(\ref{t12}),(\ref{t18}), (\ref{t27}),(\ref{t28}),
it remains to verify that the two Poisson
brackets $\pua{\vo{\Phi_0}}{\Phi_0}$ and
$\pua{\vo{{\cal H}^0}}{{\cal H}^0}$ are zero. This is a direct
calculation, which is very cumbersome, especially in the case of
the Poisson bracket $\pua{\vo{{\cal H}^0}}{{\cal H}^0}$.
We have thus proved that
constraints (\ref{t12}),(\ref{t18}),(\ref{t27}),(\ref{t28}) are first-class
constraints.

In accordance with everything said after formula
(\ref{t18}), we can write the Hamiltonian of the embedding theory with
the additionally imposed Einstein's constraints in the form of a
linear combination of all eight constraints with the Lagrange
multipliers,
\disn{t73}{
H=\int\! d^3x\ls \la^i\Phi_i+\la^0\Phi_0+N_i{\cal H}^i+N_0{\cal H}^0\rs.
\nom}
But because four of the eight constraints appear not from the
canonical formalism construction but are merely imposed
artificially and added to the Hamiltonian with their Lagrange
multipliers, the equations of motion generated by this Hamiltonian
may not exactly reproduce the initial equations of the embedding
theory and may contain some of the Lagrange multipliers as
additional variables.

The presence of the eight first-class
constraints suggests the presence of an eight-parameter gauge
symmetry in the theory, while only a four-parameter gauge group
corresponding to changing the coordinates on the surface $W^4$ is
present in the initial theory. This means that the additional
symmetry transformations must act on the additional variables (the
Lagrange multipliers).

It can be conjectured that cleverly fixing
the arising additional gauge freedom, i.~e., imposing some special
conditions on the Lagrange multipliers arising in the equations of
motion, results in these equations becoming the embedding theory
equations and correspondingly, by the satisfaction of the
constraint equations, Einstein's equations. In the next section,
we show that this is true.

It is interesting that a simple calculation shows that the quantity
\disn{t73.1}{
\pi^{lm}=-\frac{1}{2}\pi^a\al_a^{lm}
\nom}
is canonically conjugate to the three-dimensional metric $\tr{g}_{ik}$.
If we confined ourself to considering only the
dynamics of quantities composed from $\tr{g}_{ik}$ and $\pi^{lm}$,
then by virtue of formulas (\ref{t68}),(\ref{t62}), we would
drop the term in Hamiltonian (\ref{t73}) that is proportional
to $\Phi_0$ and replace $\Phi_i$ with $-{\cal H}_i$. As a result, the
Hamiltonian would become
\disn{t73.2}{
H=\int\! d^3x\ls (N_i-\la_i){\cal H}^i+N_0{\cal H}^0\rs=\ns=
\int\! d^3x\ls 2(N_i-\la_i)
\sqrt{-\tr{g}}\;\tr{\na}_k\!\ls
\frac{\pi^{ik}}{\sqrt{-\tr{g}}}\rs+2N_0\ls
\frac{\pi^{ik}\hat L_{ik,lm}\pi^{lm}}{\sqrt{-\tr{g}}}-
\sqrt{-\tr{g}}\;\tr{R}\rs\rs,
\nom}
in which form it reduces to a combination of only four constraints
rather than eight and coincides exactly with the known expression
for the Hamiltonian in the Arnowitt-Deser-Misner formalism.

\section{The action for the embedding theory with additional Einstein's
constraints}
We construct the action corresponding to Hamiltonian (\ref{t73}). For
this, we calculate the quantity $\dot y^a$:
\disn{t74}{
\dot y^a=\frac{\de H}{\de \pi_a}=
\la^i\tr{e}^a_i+\la^0w^a+\al^{a,ik}\,\tr{\na}_iN_k+
\frac{N_0}{\sqrt{-\tr{g}}}\;\al^{a,ik}\hat L_{ik,lm}\al^{lm}_b\pi^b.
\nom}
Contracting this equality with the quantity
$\tr{b}_{a,pr}$ and using its properties
and formulas (\ref{t20}), (\ref{t21}), we find the relation
\disn{t77}{
\pi^a\al_a^{ik}=\frac{\sqrt{-\tr{g}}}{N_0}\;L^{ik,lm}\ls
\dot y_a\tr{b}^a_{ik}-\frac{1}{2}\ls\tr{\na}_iN_k+\tr{\na}_kN_i\rs\rs.
\nom}
Substituting relations (\ref{t74}) and (\ref{t77})
in the formula relating the Lagrangian and Hamiltonian of the theory,
we can easily obtain the expression for the desired action,
\disn{t78}{
S=\int dx^0\ls \int d^3x\;\pi_a\dot y^a-H\rs=
\int d^4x \ls
\frac{N_0}{2\sqrt{-\tr{g}}}\;\pi^a\al_a^{ik}\hat L_{ik,lm}\al^{lm}_b\pi^b
+2N_0\sqrt{-\tr{g}}\,\tr{R}\rs=\no=
\int d^4x \sqrt{-\tr{g}} \left[
\frac{1}{2N_0}
\ls\dot y_a\tr{b}^a_{ik}-\frac{\tr{\na}_iN_k+\tr{\na}_kN_i}{2}\rs
L^{ik,lm}
\ls\dot y_b\tr{b}^b_{lm}-\frac{\tr{\na}_lN_m+\tr{\na}_mN_l}{2}\rs
+\right.\ns+2N_0\,\tr{R}\Biggr].
\nom}
We see that together with the initial theory variables $y^a$, the
Lagrange multipliers $N_i$ and $N_0$ enter this action as additional
independent variables. Hence, these variables also enter the
equations of motion obtained from action (\ref{t78}). The result is that
such equations of motion do not exactly reproduce initial
equations (\ref{nn2}) of the embedding theory.

Equations of motion
corresponding to action (\ref{t78}) impose no restrictions on the time
evolution of the variables $N_i$,$N_0$ because they are Lagrange
multipliers. These variables can be assigned arbitrarily chosen
values using the additional gauge transformations discussed above.

Comparing expression (\ref{t78}) with initial action (\ref{t1}),
we can easily
see that they coincide under the conditions
\disn{s1}{
N_i=0,\qquad
N_0=\frac{1}{2}\,\sqrt{\dot y^a\;\tr{\po}_{ab}\;\dot y^b}=
\frac{1}{2\sqrt{g^{00}}}
\nom}
(equality (\ref{t0}) is used here) on the Lagrange multipliers
$N_i$ and $N_0$. This means that if we impose conditions (\ref{s1}),
thus partially fixing the gauge freedom in the equations of motion
obtained from action (\ref{t78}), then these equations become embedding
theory equations (\ref{nn2}) supplied with Einstein's constraints (the
latter appear after varying the action over the variables $N_i$,$N_0$).
As shown above, if the initial data are in the general
position, then this set of equations is equivalent to Einstein's
equations.

We can therefore conclude that a theory equivalent to
Einstein's GR can be obtained by partially fixing the gauge
freedom in the generalized embedding theory with action (\ref{t78}),
which has the eight-parameter gauge symmetry.

It is interesting that action (\ref{t78}) up to nonintegral terms
can be written in the form of the initial GR action
\disn{s2}{
S=\int d^4x\, \sqrt{-g'}\;R(g'),
\nom}
if we here substitute for the metric $g'_{\m\n}$ not its
induced expression (\ref{nn1}) but the modification of it
\disn{s3}{
g'_{ik}=\tr{g}_{ik}=\dd_iy^a\dd_ky_a,\qquad
g'_{0k}=\dd_0y^a\dd_ky_a-N_k,\qquad
g'_{00}=4N_0^2+g'_{0i}\,\tr{g}^{ik}\,g'_{0k},
\nom}
whence we obtain $g'^{00}=\frac{1}{4N_0^2}$.
To see this, we must rewrite action (\ref{s2}) in form (\ref{77.1})
replacing $g_{\m\n}$ with $g'_{\m\n}$ and using the formula
$g'=\tr{g'}/g'^{00}$:
\disn{s4}{
S=
\int d^4x \sqrt{-\tr{g'}} \left[
\frac{1}{\sqrt{g'^{00}}}K_{ik}(g')L^{ik,lm}(g')K_{lm}(g')
+\frac{1}{\sqrt{g'^{00}}}\,\tr{R}(g')\right].
\nom}
Using the known relation
\disn{s5}{
K_{ik}=\frac{\sqrt{g^{00}}}{2}\ls
-\dd_0g_{ik}+\tr{\na}_i\,g_{0k}+\tr{\na}_k\,g_{0i}\rs,
\nom}
formulas (\ref{s3}), and the equality
\disn{s6}{
K_{ik}=\sqrt{g^{00}}\,\dot y_a\tr{b}^a_{ik},
\nom}
which follows from (\ref{n3.1b}),(\ref{n6.1}), we can see that
\disn{s7}{
K_{ik}(g')=\frac{1}{2N_0}
\ls\dot y_a\tr{b}^a_{ik}-\frac{\tr{\na}_iN_k+\tr{\na}_kN_i}{2}\rs.
\nom}
Substituting this expression in (\ref{s4}) and taking formulas
(\ref{s3}) and the fact that $L^{ik,lm}(g')\!=L^{ik,lm}(g)$ and
$\tr{R}(g')=\tr{R}(g)$ (where $g$ is the induced metric)
into account, we can easily see that the result coincides with
formula (\ref{t78}).

As stated above, the considered theory with the independent
variables $y^a$, $N_i$, $N_0$ whose action can be written in form (\ref{s2}),
has an eight-parameter gauge symmetry. We see that the quantity
$g'_{\m\n}$ is invariant under four of these eight transformations; the
generators of the former in the canonical formalism are the
constraint $\Phi_0$ and the combinations of the constraints
${\cal H}_i+\Phi_i$.
Those are the transformations that are isometric bendings of
surfaces of the constant time $W^3$ (cf. formulas (\ref{t68}),(\ref{t62})
and the reasoning after them) supplied with the corresponding
transformations of the variables $N_i$,$N_0$. We can also show that
the quantity $g'_{\m\n}$ behaves as a tensor under the remaining four
transformations, which just results in action (\ref{s2}) being invariant
under these transformations.

After gauge conditions (\ref{s1}) are
imposed, the quantity $g'_{\m\n}$ coincides with the induced metric.
Therefore, it satisfies Einstein's equations in this gauge (this
is true if the initial data are in the general position; see
above). But because this quantity is invariant under the
transformations that we use to reduce arbitrary values of
variables to those restricted by gauge conditions (\ref{s1}), it
satisfies Einstein's equations even if we do not impose gauge
conditions (\ref{s1}). Therefore,
we can in principal consider the quantity $g'_{\m\n}$
to be the metric, which is invariant under additional
symmetry transformations and coincides with the induced metric
only in gauge (\ref{s1}).

\vskip 0.5em
{\bf Acknowledgments.} The work was supported in
part by the Russian Ministry of Education
(Grant No.~RNP.2.1.1.1112), the President of the Russian
Federation (Grant No.~NS-5538.2006.2), the Russian Foundation for
Basic Research (Grant No.~05-02-17477, S.~A.~P.).

\end{document}